\documentstyle[twoside,fleqn,espcrc2,psfig]{article}

\newcommand{\AmS}{{\protect\the\textfont2
  A\kern-.1667em\lower.5ex\hbox{M}\kern-.125emS}}

\def\arcdeg{\hbox{$^\circ$}}

\title{Outbursts, State Transitions, and Periodicities Observed with
the RXTE All-Sky Monitor}

\author{Alan M. Levine\address{Center for Space Research,
        Massachusetts Institute of Technology, Cambridge, MA 02139,
        USA}}
       
\begin{document}

\begin{abstract}
Results from the All-Sky Monitor (ASM) on the {\it Rossi X-ray Timing
Explorer} are reviewed.  A number of recurrent transient sources have
been detected, while only a few previously unreported sources have
been discovered.  The ASM light curves show a wide variety of phenomena
in general, and, in particular, those of transient sources show a wide
range of properties. Examples are used to illustrate that the
distinction between persistent and transient sources may be very
unclear.  The results of searches for periodicities in the ASM light
curves are summarized, and other astrophysical investigations using
ASM light curves are suggested.  The latter include investigations of
the possible causes of long-term quasiperiodic and chaotic
variability, and comparative studies on the basis of the observed
variability.
\end{abstract}

\maketitle

\section{Introduction}

The primary purpose of the All-Sky Monitor on the {\it Rossi X-ray
Timing Explorer} is to alert the astrophysics community in a timely
manner to the outbursts of transient X-ray sources and to changes of
state of persistent sources.  Another purpose, which some may be
consider to be a byproduct, is to record long-term X-ray ``light
curves'' for the detected sources.  Here I review the results of a year
and a half of operation of the ASM, with selected illustrative
examples of light curves and other results.  First, the instrument
design, operation, and data analysis is briefly described to provide
background useful in interpreting the results.  The numbers and types
of transient sources that have been detected so far by the ASM are
then discussed.  A selection of astrophysical investigations that may
be supported by the ASM results is described.  Finally, I give advice
and encouragement on the use of ASM data and results.

\section{Instrumentation and Operations}
 
The ASM consists of three Scanning Shadow Cameras (SSCs) mounted on a
motorized rotation drive. Each SSC contains a position-sensitive
proportional counter (PSPC) that views the sky through a slit
mask. The PSPC is used to measure the displacements and strengths of
the shadow patterns cast by X-ray sources within the field of view,
and to thereby infer the directions and intensities of the
sources. Each SSC is sensitive in the energy range of approximately
1.5--12 keV, with on-axis effective areas of $\sim$10 cm$^2$, $\sim$30
cm$^2$, and $\sim$23 cm$^2$ at 2, 5, and 10 keV, respectively, and
with a $6\arcdeg \times 90\arcdeg$ (FWHM) field of view.

The ASM is operated so as to obtain series of 90--s exposures, called
``dwells'', during which the orientations of the SSCs remain fixed.
The drive assembly rotates the structure holding the 3 SSCs by
$6\arcdeg$ between dwells so that large portions of the sky are
eventually scanned during each orbit of the spacecraft.  The
sensitivity is typically $\sim$30 mCrab ($3 \sigma$) for a source in
the central region of the field of view of an SSC for a 90--s
exposure.  Any given source is typically scanned 5 to 15 times over
the course of a day, so that the sensitivity is of order $\sim$10
mCrab on a daily basis.

A more extensive description of the instrument, operations, and data
analysis may be found in Levine et al.~\cite{Levi96}.

\section{Transient and Newly Discovered Sources}

\begin{figure*}
\centerline{\psfig{figure=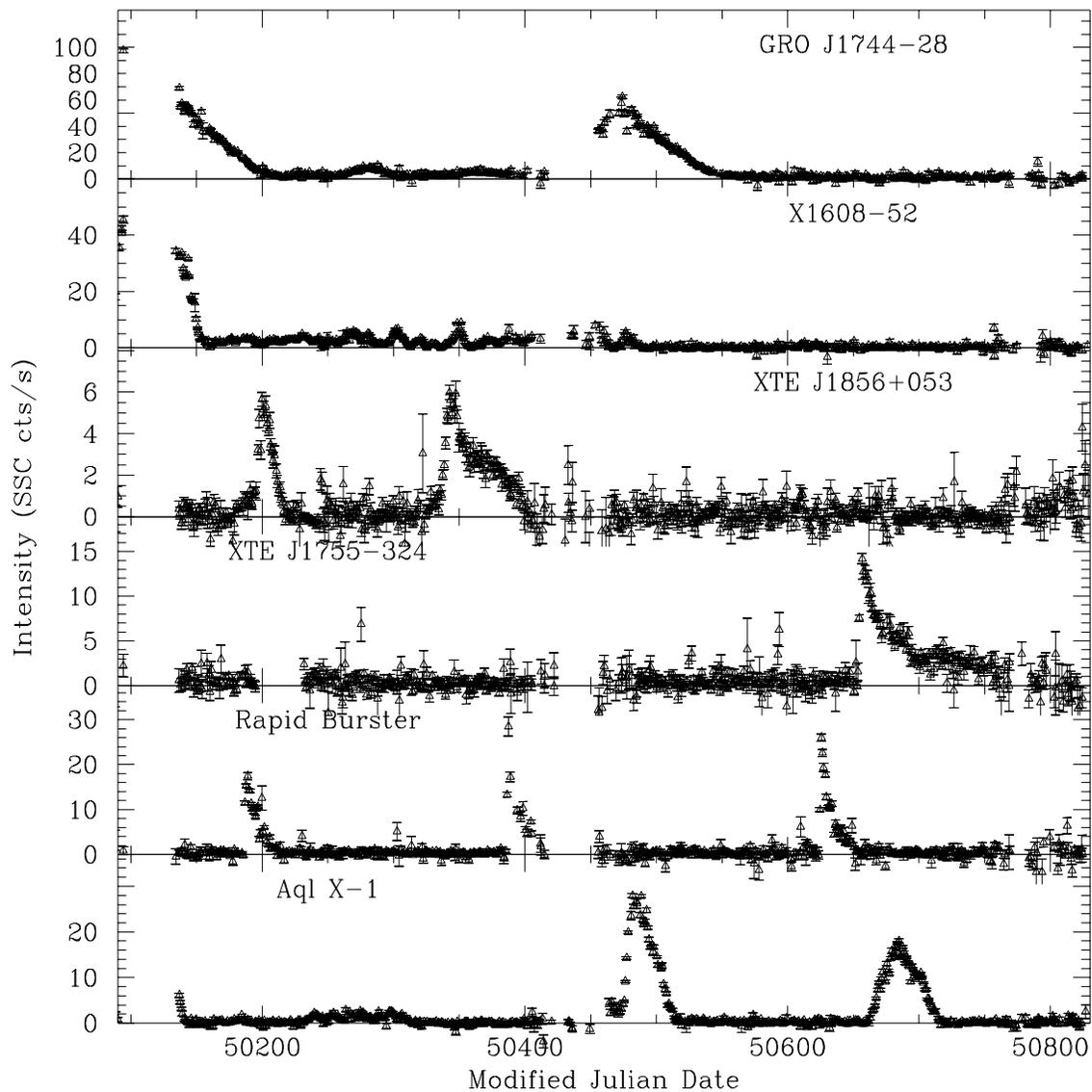,width=16cm,height=17cm} }
\caption{ASM light curves (1.5--12 keV) of 6 sources showing
transient outbursts.  Each data point is the weighted average
intensity of all of the intensity measurements obtained from 90--s
observations taken on a given day.  The intensity is given in units of
the count rate corrected for location in the field of view and for
differences among the 3 SSCs.  The Crab Nebula, for reference, yields
75 cts/s. Modified Julian Date 50083 corresponds to 1996 January 1.}
\label{fig:romefig1}
\end{figure*}

When the ASM first became operational, two sources were in outburst
and were bright enough to be easily detectable.  These outbursts were
associated with the sources GRO J1744--28 and X1608--52 (see
Figure~\ref{fig:romefig1}).  The ASM has since detected the outbursts
of quite a few other sources.  While the majority of these sources
were previously known or were discovered by other means, three of these
transient sources were newly discovered with the PCA, and two were newly
discovered using the ASM.  This small number of new transient sources
was a surprise. Since the ASM is significantly more sensitive than
previous X-ray monitors, we anticipated that more previously unknown
sources would be found.  The sources that have been discovered with
the PCA include XTE J1856+053, XTE J1842--042, and XTE J1739--302, while
XTE J1716--389 and XTE J1755--324 were discovered with the ASM.

The light curves of 6 selected sources which have been detected in
outburst for limited periods of time are shown in
Figure~\ref{fig:romefig1}.  As noted above, the bursting pulsar GRO
J1744--28 and the LMXB X1608--52 were already in outburst at the
beginning of the {\it RXTE} mission.  The middle two light curves are
of sources discovered with {\it RXTE}; one can see that XTE J1856+053
underwent two distinct weak outbursts, while only one outburst of XTE
J1755--324 has been observed.  The spectrum of XTE J1755--324 was very
soft near the peak of its outburst, and in fact was softer than any
other source in the galactic center region that can be detected with
the ASM. Its spectrum hardened considerably as its intensity decayed.
The light curve of the Rapid Burster (X1730--333) shows recurrent
outbursts which are spaced by $\sim$200 and $\sim$240 days.  This
confirms earlier indications that the source tends to go into outburst
about every six months \cite{Lewi95}, and indicates that the outbursts
most likely occur quasiperiodically.  The bottom panel of
Fig.~\ref{fig:romefig1} shows the outburst behavior of Aql X-1 in the
{\it RXTE} era.  In addition to the two obvious outbursts, it appears
that the ASM just caught the tail end of an outburst as it first
became operational, and detected a ``failed outburst'' starting around
MJD 50240.

\begin{figure*}
\centerline{\psfig{figure=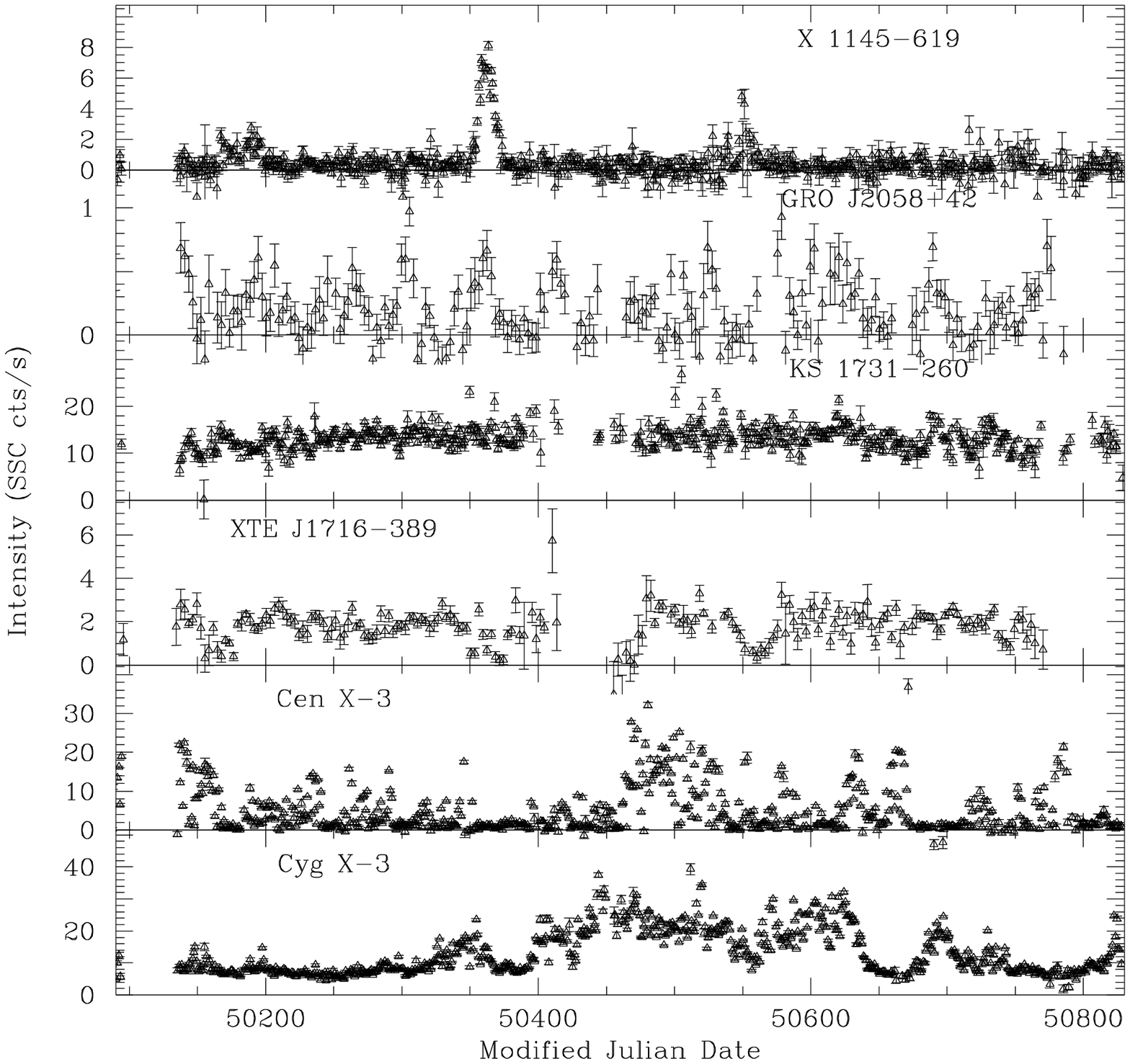,width=16cm,height=17cm} }
\caption{ASM light curves (1.5--12 keV) of 6 more sources.  The light
curves for GRO J2058+42 and XTE J1716--389 comprise 3--day average
intensities; the other light curves comprise 1--day averages.}
\label{fig:romefig2}
\end{figure*}

A number of systems show outbursts which appear according to an
underlying periodicity. For systems such as GX301--2, EXO2030+375, and
GRO J2058+42, the outbursts are consistently detected, whereas in
systems like 4U1145--619 and 4U0115+63 only a few outbursts have been
detected.  In the latter two sources, the outbursts are spaced at 187
d and 24 d, respectively.  I should note that in GRO J2058+42 the
outbursts occur every $\sim$55 d, while a spacing of $\sim$110 d is
apparent in the light curve derived from observations with
BATSE~\cite{Wils98}.  The light curves of 4U1145--619 and GRO J2058+42
may be seen in Fig.~\ref{fig:romefig2}.

Some sources, such as GRS1915+105 and GRO J1655--44, have undergone
relatively long-lived outbursts.  Other sources, such as KS1731--260
and XTE J1716--389 (Fig.~\ref{fig:romefig2}), would not be classified
as transients on the basis of {\it RXTE} data alone, since their emission
has persisted for the entire mission so far, but are nonetheless
classified as transients since they were not detected in observations
with other instruments on occasions prior to the launch of {\it RXTE}.

We further illustrate the muddiness of the distinction between
transient and persistent sources with the light curves of Cen X-3 and
Cyg X-3 (Fig.~\ref{fig:romefig2}).  One can see the large and
unpredictable variation of the intensities of these two sources, which
is not at all unprecedented among the so-called ``persistent'' sources.
Persistent sources also show state changes, with the hard/soft
transitions of Cyg X-1 being the archetype~\cite{Zhan97a}.

The ASM has also detected X-ray emission during gamma-ray bursts and
provided data for burst localization.  Early estimates of the numbers
of gamma-ray bursts were low becuse of a lack of data on the X-ray
intensity of bursts and because the long duration, often over 100 s,
of bursts in the 2--12 keV band was not taken into account.  Since
the ASM typically rotates by $6\arcdeg$ every 90 s, the effective sky
coverage for events that last $\sim$100 s is of order twice as large
as for events which last a small fraction of 90 s.  Indeed, this
factor made it possible for the ASM to obtain crossed lines of
position from the 1997 August 15 and 1997 August 28 gamma-ray bursts.

\section{Astrophysics from ASM Light Curves}

In addition to alerting observers to interesting cosmic X-ray
phenomena, the ASM light curves may be directly of use in
astrophysical investigations.  One of the most obvious of these is a
search for periodic variability that could reveal the orbital period
of a system or some other interesting phenomenon.  Another type of
investigation is the comparative study of the variability of different
sources.  In this section, I suggest a few lines of investigation that
are likely to be interesting.

\begin{figure*}[!htp]
\centerline{\psfig{figure=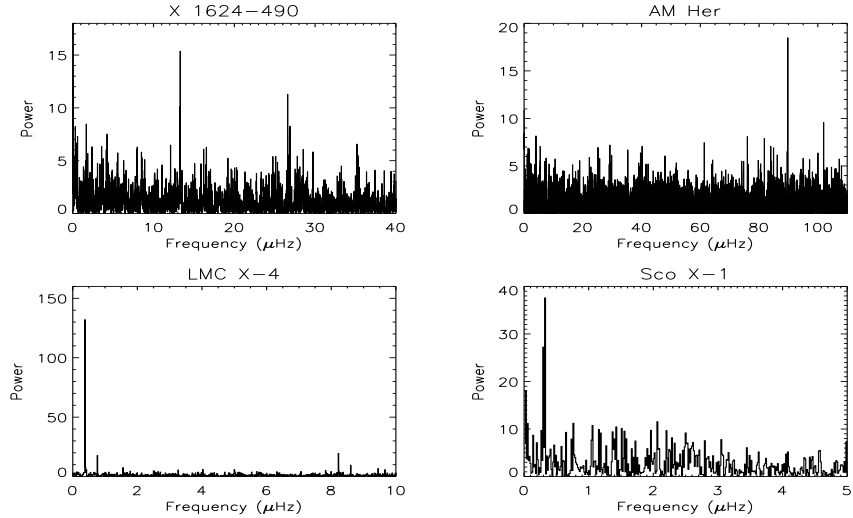,width=12cm,height=7cm} }
\caption{Power density spectra obtained from ASM light curves.  The
spectra are normalized to have an average power of unity.}
\label{fig:asmpds}
\end{figure*}

We and others have searched for periodic signals in the the ASM X-ray
light curves with the use of FFTs and/or Lomb-Scargle transforms.
Figure~\ref{fig:asmpds} shows examples of power density spectra
constructed using FFTs.  Significant signals are detected in the
spectra of each of the four power density spectra.  In the cases of
X1624--490, AM Her, and LMC X-4, the periodicities were known prior to
the {\it RXTE} era.  The detection of the 3.09 h periodicity of AM Her
was unexpected, in that the source is so weak ($\sim$1 mCrab) that it
is not detectable from direct inspection of the light curve itself. In
the case of Sco X-1, two peaks at periods near 37 days, stand out
(also see ~\cite{Peel96}).

Table \ref{tab:asmpers} includes a list of periodicities that have either been
detected in our analyses of ASM data or have been reported in the
literature based on ASM data.  Most of these are unquestionably
detected in the power density spectra that we have computed; a few
appear, in our analyses, to be based on peaks of marginal significance
when the generally high noise level at low frequencies is taken into
account.  The 2.6 d periodicity in GRO J1655--40 was reported by
Kuulkers et al.~\cite{Kuul97} not on the basis of a feature in a power
density spectrum, but on the basis of the detection of ``dips'' that
occur at about the same orbital phase.

\begin{table*}[hbt]
\caption{Periodicities detected with the ASM}
\label{tab:asmpers}
\begin{tabular*}{\textwidth}{@{}l@{\extracolsep{\fill}}rcl}
\hline
Source&  Period&  Origin&  Reference \cr
\hline
AM Her&  3.094 h&  orbit& \cr
Cen X-3&  2.083 d&  orbit& \cr 
Cir X-1&  16.6 d&  orbit (?)& \cr
Cyg X-3&  4.78 h&  orbit (?)& \cr
EXO2030+375& 46 d&  orbit& \cr
GX13+1&  24.7 d&  & Corbet \cite{Corb96c}\cr
GX301--2& 41.5 d&  orbit& \cr
Her X-1&  1.7 d, 35 d&  orbit, disk precession& \cr
LMC X-4&  1.4 d, 30 d&  orbit, disk precession& \cr
SMC X-1&  3.89 d, $\sim$60 d&  orbit, disk precession& Zhang et al. \cite{Zhan96}, Wojdowski et al. \cite{Wojd98} \cr
Vela X-1& 8.96 d&  orbit& \cr
X 0114+65& 2.74 h, 11.7 d&  n.s. rotation, orbit&  Corbet \& Finley \cite{Corb96d}\cr
4U1538--52& 3.73 d&  orbit& \cr
4U1700--37& 3.41 d&  orbit& \cr
X1822-37& 5.57 h&  orbit& \cr
Cyg X-2&  77 d&  superorbital& Wijnands, Kuulkers, \& Smale \cite{Wijn96} \cr
GRO J2058+42& 54 d&  1/2 orbit (?)& Corbet, Peele, \& Remillard \cite{Corb97b}\cr
Sco X-1&  37 d&  superorbital& Peele \& White \cite{Peel96}\cr
X0115+63& 24.3 d&  orbit& \cr
X1907+097& 8.38 d&  orbit& \cr
X1657+415& 10.4 d&  orbit& \cr
4U1145--619& 187 d&   & Corbet \& Remillard \cite{Corb96b}\cr
Cyg X-1&  5.6 d&  orbit&  Zhang, Robinson, \& Cui \cite{Zhan96b}\cr
X0726-260& 34.5 d&   & Corbet \& Peele \cite{Corb97a}\cr
GRO J1655--40& 2.621 d& orbit&  Kuulkers et al. \cite{Kuul97}\cr
X2127+119& 17.1 h, 37 d&  orbit, superorbital&  Corbet, Peele, \& Smith \cite{Corb96a}\cr
X1624--490& 21 h& orbit& \cr
X Per&  835 s&  n.s. rotation& \cr
\hline 
\end{tabular*}
\end{table*}

\begin{figure*}[hbt]
\centerline{\psfig{figure=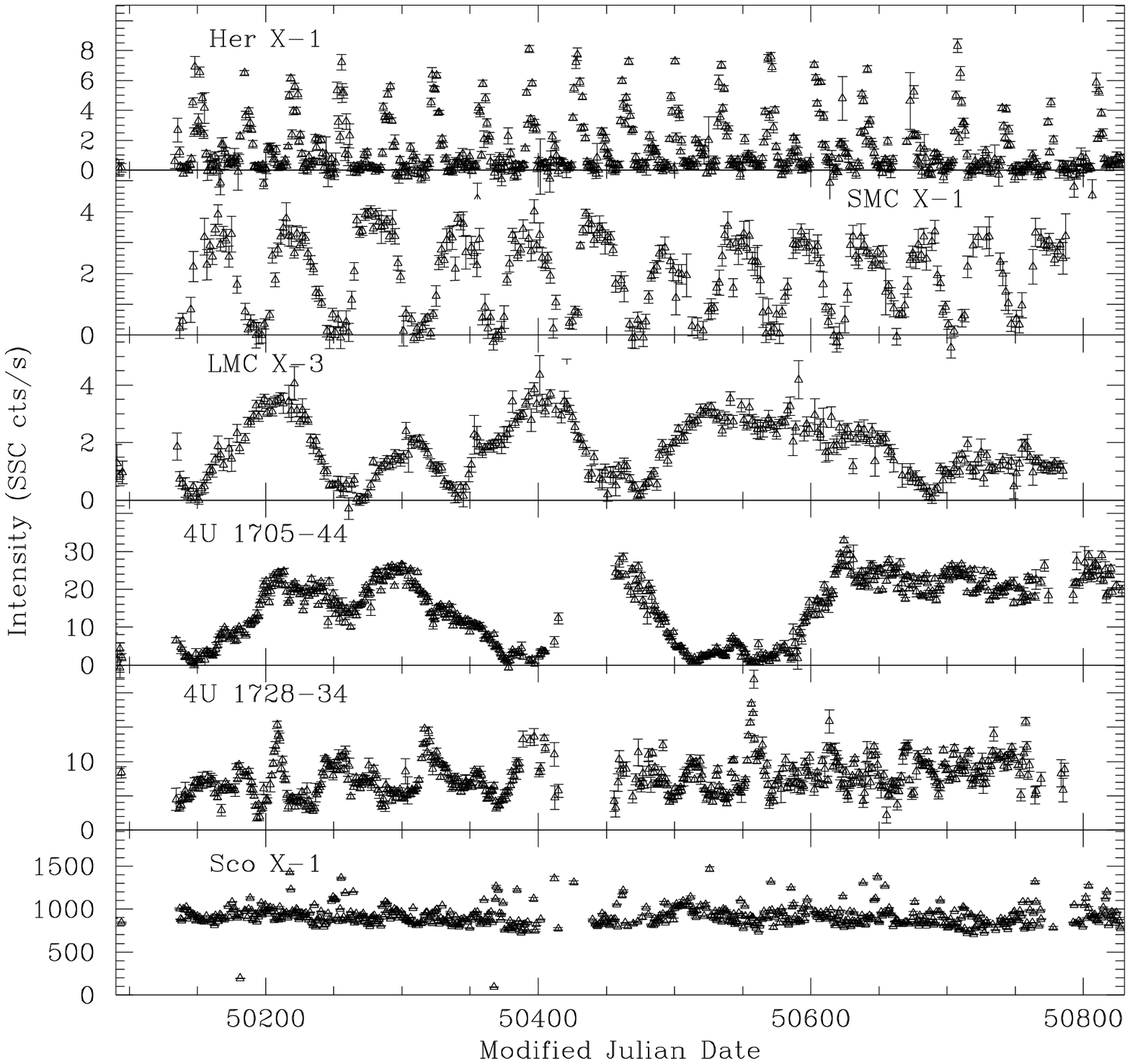,width=16cm,height=17cm} }
\caption{ASM light curves (1.5--12 keV) for 6 more sources.  For SMC
X-1, only non-eclipse data were used to construct the light curve shown
here. The light curves for SMC X-1 and LMC X-3 comprise 2--day average
intensities; the other light curves comprise 1--day averages.}
\label{fig:romefig3}
\end{figure*}

The quasiperiodic intensity variations seen in systems like Her X-1
and LMC X-4 are most likely caused by a tilted precessing accretion
disk.  The ASM light curve of SMC X-1 (Fig.~\ref{fig:romefig3})
strongly suggests that such a disk is present in this system, and this
is confirmed by observations with other instruments ~\cite{Wojd98}.
Quasiperiodic variations, on time scales much longer than the orbital
periods, have also been seen in other systems, such as Cyg X-2, LMC
X-3, and 4U1705--44 (see Fig.~\ref{fig:romefig3}).  It is natural to
raise the question of whether the quasiperiodic variability in these
sources is also caused by tilted (or warped) precessing accretion
disks.  This question can be extended to asking whether the apparently
random variability of 4U1728-34 (see Fig. 3), or the variability of
4U1820-30, which seems to have a periodicity near P = 175 d (Smale \&
Lochner~\cite{Smal92}) as well as random variability that is similar
to that of 4U1728-34, may also be caused by the same effect.  One can
speculate that disk tilt, warp, and/or precession may be evident as
variability that can range from being rather periodic to being quite
chaotic depending on the details of the accretion rate and disk
structure.

The light curves of the Z-type LMXB's show definite similarities to
each other.  In particular, the light curves of the three Z sources
Sco X-1 (see Fig.~\ref{fig:romefig3}), GX17+2, and GX349+2, show
flares on top of a relatively constant baseline.  The Z sources GX5--1
and GX340+0 yield light curves which are better described by random
variability within a limited intensity range rather than by a baseline
with distinct flares.  Finally, Cyg X-2 shows dips from a baseline
that wanders in a quasiperiodic manner over a range of a factor of 3
or more in strength.  From Figure~\ref{fig:romefig3}, one can see that
the light curve of Sco X-1 bears no resemblance at all to that of
atoll sources like 4U1728--34.

Another result that begs for explanation is the large degree of
variation of the profiles and durations of the outbursts of transient
sources.  The outbursts of 4U1630--47, GRS1915+105, GRO J1655--40, and
GRS1739--278 are each quite different from any of the outbursts shown
in Fig.~\ref{fig:romefig1}.  I can also mention that the light curves
and periodic spectral variation of Cir X-1 must be telling us
something of astrophysical importance, and dramatic spectral
variations of Cyg X-1 likewise.

\section{Conclusions}

Observers and students of bright X-ray sources will find ASM data and
results to be useful.  The light curves can indicate good observing
opportunities, and later, when interpretation is in order, can give
context to ``snapshot'' observations.  One can keep informed using the
MIT and GSFC web pages (http://space.mit.edu/XTE and
http://heasarc.gsfc.nasa.gov/docs/xte/XTE.html) which have notes on source
activities as well as the light curves.  Detailed analyses of a number
of the brighter sources will be interesting; the data are available to
be downloaded and analyzed.  These data include the individual
intensity measurements from the 90 s dwells in each of three energy
bands.

While I strongly encourage the use of ASM data, I should caution
potential users to beware of the occasional individual measurement
that is affected by a problem.  Patterns in the results are often much
more reliable than a single high or low measurement.  Finally, our
error estimates do not include contributions representing all
systematic effects.  The errors are undoubtedly underestimated for
sources observed with other sources also present in the field of view.
The systematic effects will be particularly obvious in long-term
averages of measurements of weak sources.  I would also encourage
potential users to contact a member of the ASM team with any questions
that may arise.

The results from the ASM have been obtained by the direct efforts of
many individuals at MIT and GSFC and by the indirect efforts of many
others.  The author is grateful for all their contributions.

\end{document}